\documentclass{article}
\usepackage{spconf,amsmath}
\usepackage[pdftex]{graphicx}
\usepackage{amsmath}

\title{SPEAKER SELECTIVE BEAMFORMER WITH KEYWORD MASK ESTIMATION}
\name{Yusuke Kida, Dung Tran, Motoi Omachi, Toru Taniguchi, Yuya Fujita}
\address{Yahoo Japan Corporation}
\begin{document}
%
\maketitle
\begin{abstract}
This paper addresses the problem of automatic speech recognition (ASR) of a target speaker in background speech.
The novelty of our approach is that we focus on a wakeup keyword, which is usually used for activating ASR systems like smart speakers.
The proposed method firstly utilizes a DNN-based mask estimator to separate the mixture signal into the keyword signal uttered by the target speaker and the remaining background speech.
Then the separated signals are used for calculating a beamforming filter to enhance the subsequent utterances from the target speaker.
Experimental evaluations show that the trained DNN-based mask can selectively separate the keyword and background speech from the mixture signal.
The effectiveness of the proposed method is also verified with Japanese ASR experiments, and we confirm that the character error rates are significantly improved by the proposed method for both simulated and real recorded test sets.
\end{abstract}
\renewcommand{\thefootnote}{\fnsymbol{footnote}}
\footnote[0]{\copyright 2018 IEEE}

\section{Introduction}
Robustness against background speech is one of the key factors for automatic speech recognition (ASR).
Even if multiple talkers are surrounding us and speaking simultaneously, we can focus on a specific target speaker.
This is called the {\it cocktail-party effect} \cite{Moray}, and this function is realized by our auditory system.
On the other hand, current ASR systems cannot handle such a situation and therefore the background speech usually causes a serious performance degradation in ASR.
One practical example of this scenario is a smart speaker located in a living room.
In this situation, background speech from surrounding people, television and radio are expected to overlap the target speech.

A simple way to realize robust ASR against background speech is to introduce blind speech separation processing before recognition.
Blind speech separation has been studied for decades, and it can be divided into single-channel and multi-channel approaches.
The single-channel approach is based on the spectral characteristics of each source signal.
Non-negative matrix factorization (NMF) \cite{smaragdis} and time-frequency masking \cite{Yilmaz} are well known methods.
In addition, recent deep learning-based technologies such as permutation invariant training (PIT) \cite{Yu} or deep clustering \cite{Hershey} have caught a great deal of researchers' attention.
The multi-channel approach is based on the spatial information.
Independent component analysis (ICA) \cite{Smaragdis1998blind}\cite{Kim}, full-rank spatial covariance model-based methods \cite{Duong}\cite{Sawada} are typical solutions, and also deep learning-based techniques have been studied recently \cite{Yoshioka}.

The mixture signal of a target speech with background speech can be separated into individual source signals by the above blind speech separation algorithms.
However, these algorithms cannot identify which output signal corresponds to the target speech to be recognized.
This problem is regarded as one of the permutation problem \cite{Delcroix}, and there has been some work to try to solve it by imposing constraints about speaker gender \cite{Weng} or signal intensity \cite{Wang}.
However, such constraints are not necessarily satisfied.
\^{Z}mol\'{i}kov\'{a}, et al. proposed a method named SpeakerBeam, which extracts target speech directly without any constraints about signal characteristics \cite{Zmolikova}\cite{Delcroix}.
Their method instead assumes that the target speaker is known in advance, and requires a pre-recorded clean utterance from the target speaker.

In this work, we propose an alternative approach to realize robust ASR of a target speaker in background speech while avoiding the permutation problem.
The novelty of our method is that we focus on a wakeup keyword like "okay Google" or "alexa", which is used for activating ASR systems such as a smart speaker.
These systems usually assume that the target speaker speaks a specific keyword and then a command to be recognized.
Therefore, it is naturally considered that the keyword utterance provides some important cues about the target speaker, which is beneficial for recognizing a subsequent command utterance.
Motivated by this, the proposed method utilizes the keyword utterance to estimate spatial information of the target speaker.
From the same point of view, King, et al. also proposed to utilize the keyword to calculate the mean value for feature normalization used for the acoustic modeling \cite{King}.
Our proposed method firstly separates the mixture signal into the keyword and the remaining background speech using a specially designed DNN-based mask estimator.
Then the separated signals are used for calculating a beamforming filter to enhance the subsequent utterances from the target speaker.
An advantage of the proposed method compared to SpeakerBeam is that the proposed method can enhance any speaker's utterance where the keyword is spoken, and it does not require the pre-recording procedure.

Signal-level and ASR evaluations are performed to verify the effectiveness of our proposed method.
We use two Japanese test sets for the evaluation.
The first test set is composed of simulated mixture signals between two speakers, and the second set is of realistic utterances recorded under television sound.

The rest of the paper is structured as follows.
Sections 2 and 3 describe the details of our proposed method.
The training and test sets used are described in Section 4.
The experimental evaluations and results are shown in Sections 5 and 6.
Section 7 presents our conclusions and future work.

\section{DNN-based keyword mask estimation}
\subsection{Network configuration}
\begin{figure}[tb]
\begin{center}
\includegraphics[width=8.0cm]{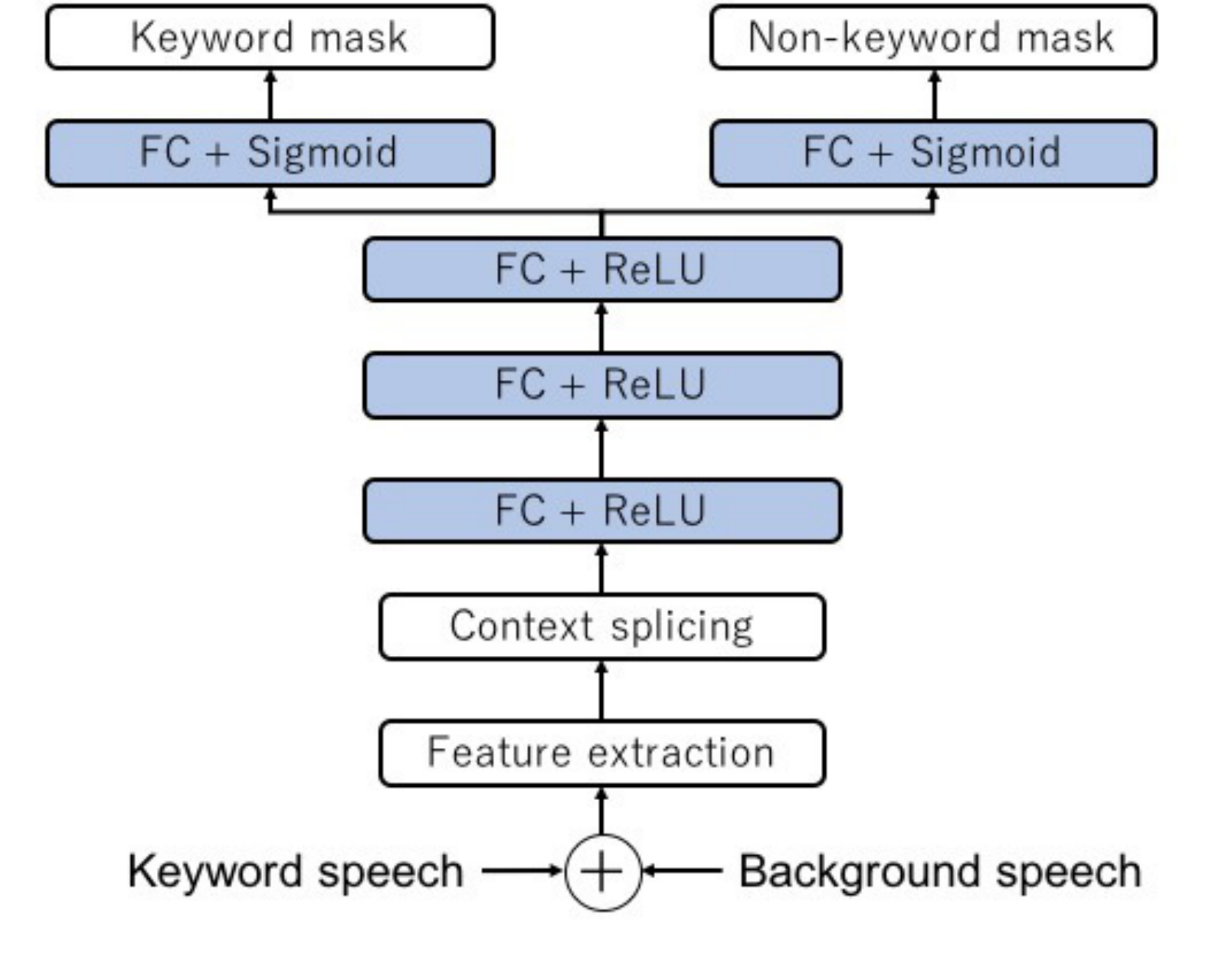}
\vspace{-10pt}     
\end{center}
\caption{Our DNN structure for keyword mask estimation. The output of DNN is two types of mask correspond to the keyword and the remaining non-keyword component. Colored blocks include trainable neural network parameters.}
\label{fig:1}
\end{figure}

Our DNN structure is shown in Fig.~\ref{fig:1}.
Given a mixture signal of a keyword with background speech, this DNN outputs two kinds of mask.
The first mask is a {\it keyword mask} which works as an extractor of the keyword.
The other mask is a {\it non-keyword mask} which works as a remover of the keyword, and accordingly extracts the remaining background speech.
The interested signal for the extraction or removal is always the specific keyword.
Therefore, this DNN can be trained efficiently, because the acoustic variation to be considered is far less than those used in standard speech separation problems.

The mixed signal is firstly converted to feature vectors.
In this work, magnitude spectra are used for the features.
Conditions for speech analysis are shown in Table \ref{tbl:condition}.
A following context splicing block extends the 256 dimensional magnitude spectra with its neighboring 20 context frames (from left 10 frames to right 10 frames), resulting in 5,376 dimensions.
Then mean and variance normalization is performed with its global value calculated from the entire training data.
Finally, the normalized feature vectors are fed into the DNN.

Our DNN has 3 fully connected (FC) hidden layers.
Each hidden layer has 1,024 nodes, and the output layer has 256 nodes for each of the two masks.
The activation function used in the hidden layers is a rectified linear unit (ReLU) function, but the sigmoid function is used for the output layer in order to limit the range of DNN output between 0 and 1.

\subsection{Parameter optimization}
The network parameters in the DNN are trained to minimize the error between the two output masks and a given reference.
In this work, we use an ideal binary mask (IBM) for the reference like \cite{Heymann}, and a cross entropy function is adopted for the error criterion.
The mini-batch size for the stochastic gradient descent (SGD) algorithm is set to 128.
Dropout is also used, in which the dropout rate is 0.2 for the input layer and 0.5 for each hidden layer.
The learning rate is set to 0.01.
The number of training epochs is 50.

\begin{table}
\caption{Condition for speech analysis.}
\label{tbl:condition}
\begin{center}
\vspace{10pt}   
\begin{tabular}{|l|l|}\hline
Sampling frequency & 16 kHz\\
Window type        & Hanning\\
Frame length       & 32 ms\\
Frame shift        & 16 ms\\ \hline
\end{tabular}
\end{center}
\end{table}

\section{Proposed system}
\subsection{Overview}
\begin{figure}[tb]
\begin{center}
\includegraphics[width=7.0cm]{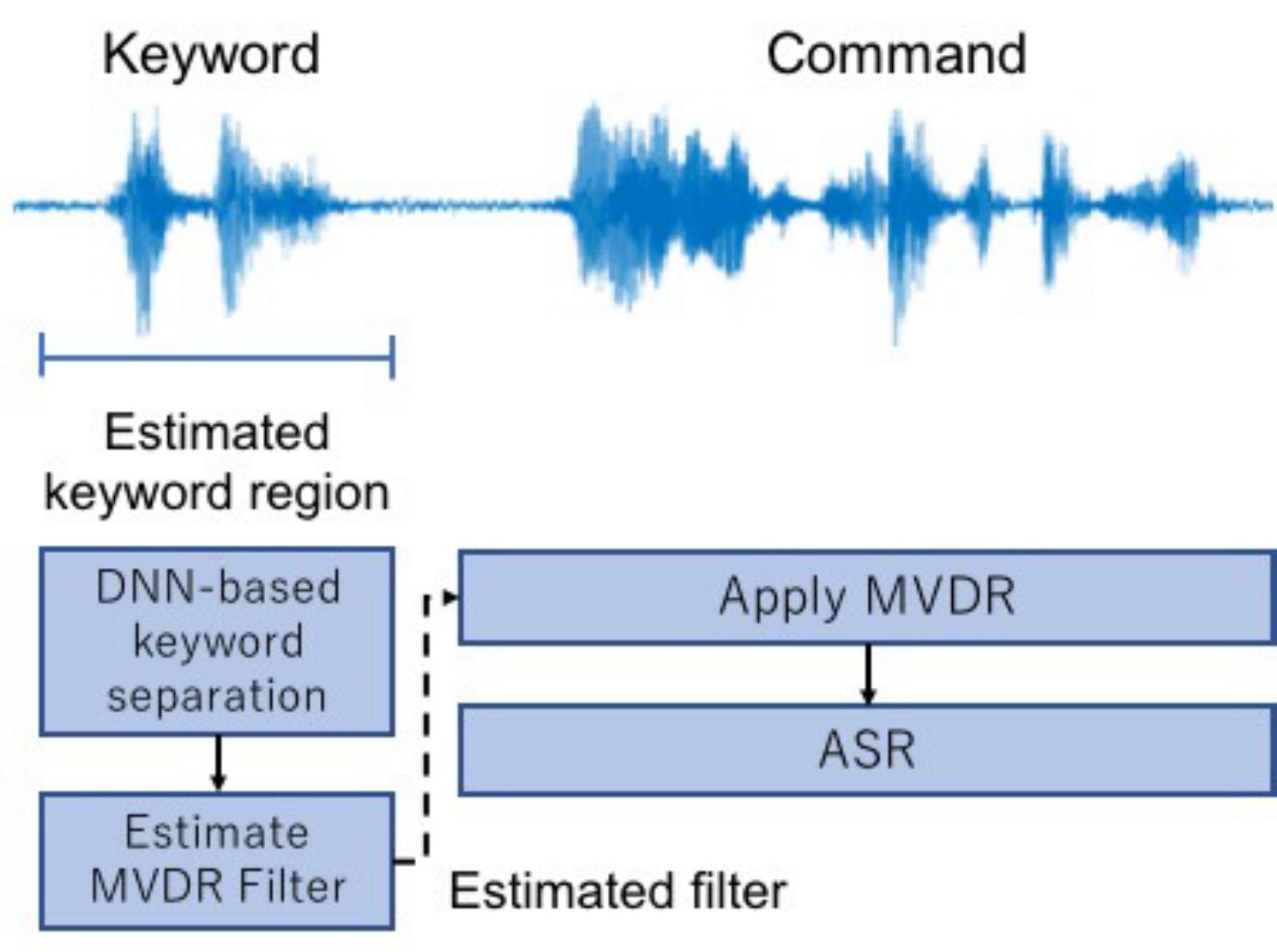}
\vspace{-10pt}     
\end{center}
\caption{A schematic diagram of the proposed system. Note that the MVDR filter is estimated during the short keyword region only and fixed.}
\label{fig:2}
\end{figure}

A schematic diagram of the proposed system is presented in Fig.~\ref{fig:2}.
The proposed system is activated when the existence of a keyword utterance is given by a keyword detection method, which is defined outside of the system.
Given a detected keyword with its estimated time region, the observed signal among the keyword region is fed into the trained DNN shown in Fig.~\ref{fig:1}.
Then the keyword and the remaining background speech are separated by applying the obtained {\it keyword mask} and {\it non-keyword mask} to the original mixture signal respectively.
This process is repeated for each of the microphone channels, and the separated multi-channel signals are then used for calculating a beamforming filter.
A well known minimum variance distortionless response (MVDR) beamformer is employed in this work.
After that, the beamforming process is applied to the subsequent signal in order to enhance target speaker utterances while reducing background speech.
Note that the beamforming filter is calculated once after the keyword, and not updated during subsequent command.
Finally, the enhanced target speaker utterance is input to the ASR system.

\subsection{MVDR filter estimation}
MVDR filter estimation can be explained as follows.
Given a background non-keyword speech covariance matrix ${\bf R}_{nn}$ and a steering vector ${\bf v}$ of keyword speech, the MVDR filter ${\boldsymbol \gamma}$ can be calculated by the following equation:
\begin{equation}
{\boldsymbol \gamma}=[\gamma(1), \gamma(2), \cdots, \gamma(c)]^t=\frac{{\bf R}_{nn}^{-1} {\bf v} }{ {\bf v}^h {\bf R}_{nn}^{-1}{\bf v}},
\label{eq:mvdr}
\end{equation}
where $c$ denotes the number of microphone, and $\cdot^t$ and $\cdot^h$ indicate transpose and conjugate transpose of a matrix, respectively.
Note that the frequency index is omitted from the above equation and the following discussion if not necessary.
In this work, ${\bf R}_{nn}$ is estimated using the observed multichannel magnitude spectra ${\bf Y}_\tau = [y_\tau(1), y_\tau(2), \cdots, y_\tau(c) ]^t$ and the {\it non-keyword mask} ${\bar m}_\tau^{(n)}$ with $\tau$ denoting the time frame index.
${\bar m}_\tau^{(n)}$ is defined as the median of ${\bf M}_\tau^{(n)}=\{m_\tau^{(n)}(1), m_\tau^{(n)}(2), \cdots, m_\tau^{(n)}(c)\}$, with $m_\tau^{(n)}(\cdot)$ denoting the estimated {\it non-keyword mask}.
Based on these, ${\bf R}_{nn}$ is calculated like \cite{Yoshioka} as:
\begin{equation}
{\bf R}_{nn} = \sum_{\tau\in {\bf T}} {\bar m}_\tau^{(n)} {\bf Y}_\tau ({\bar m}_\tau^{(n)}{\bf Y}_\tau)^h,
\label{eq:covariance-n}
\end{equation}
where ${\bf T}$ indicates the set of time frames indices among the keyword region.

The steering vector ${\bf v}=[v(1), v(2), \cdots, v(c)]^t$ can be estimated by a covariance matrix ${\bf R}_{kk}$, which can be calculated similarly as ${\bf R}_{nn}$ with the {\it keyword mask} ${\bar m}_\tau^{(k)}$:
\begin{equation}
{\bf R}_{kk} = \sum_{\tau\in {\bf T}} {\bar m}_\tau^{(k)} {\bf Y}_\tau ({\bar m}_\tau^{(k)}{\bf Y}_\tau)^h,
\label{eq:covariance-k}
\end{equation}
An eigen value decomposition of ${\bf R}_{kk}$ is calculated, and ${\bf v}$ is estimated as the eigen-vector with the maximum eigenvalue.

The estimated beamforming filter ${\boldsymbol \gamma}$ is then applied to the subsequent mixture signal as $x_\tau = {\boldsymbol \gamma}^h {\bf Y}_\tau$, and the enhanced target signal $x_\tau$ is obtained.

%
\begin{figure}[tb]
\begin{center}
\vspace{-10pt}   
\includegraphics[width=6.5cm]{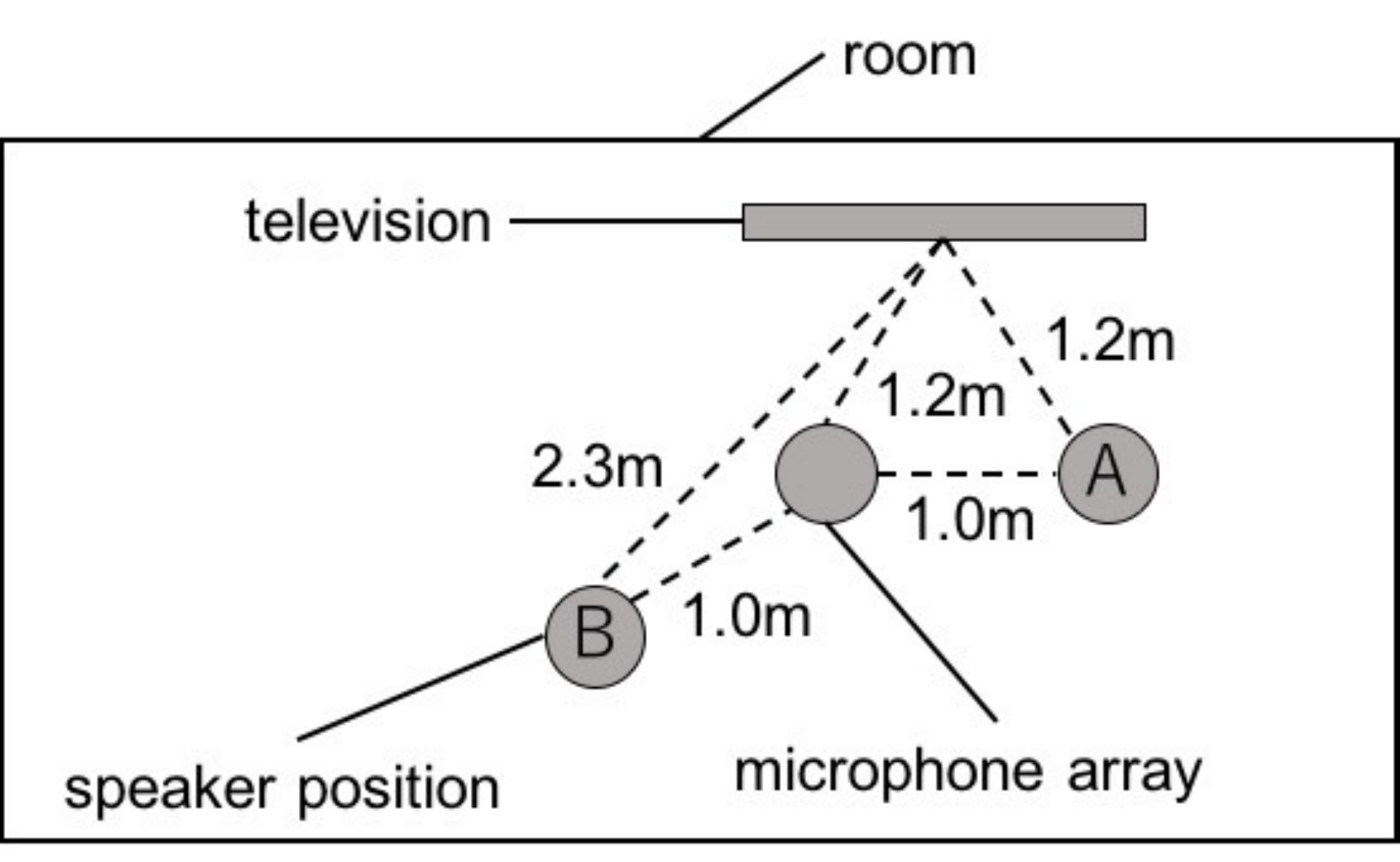}
\vspace{-15pt}     
\end{center}
\caption{Recording setting for the real-set. One speaker was located at A or B, and spoke toward the microphone array.}
\label{fig:room}
\end{figure}
\section{Data}
\subsection{Training data for keyword mask estimator}
In this work, we used our in-house Japanese data for the training and test set.
Keyword and background speech were recorded independently in the same quiet room.
We used a 4-channel microphone array for the recording.
For the training data, the recorded 4-channel signal was treated as four single-channel signals.
Speakers were located at several points in the room, and the distances between the microphone array and the speakers were between 1 and 3 meters.
The number of recorded keyword utterances was 1,660 from 35 speakers, and that of background speech utterances was 1,400 from 25 speakers.
Various combination pairs between keyword and background speech made a total of 116,200 mixed utterances, and these were used as the training set of our DNN-based mask estimator.
Note that the keyword and background speech for the mixing were selected from different speakers and their genders may be the same.
The average signal-to-noise ratio (SNR) for the mixing was 3.2 dB, and the standard deviation was 3.4 dB.
The keyword used in this work was a single Japanese word composed of 3 syllables with average duration 0.7 second.

\subsection{Evaluation data}
\label{sec:eval-data}
Two test sets were used for verifying the effectiveness of the proposed method.
They were recorded in the same room as the training data, and the same microphone array was used.
The first set was recorded in the same manner as the training set using different speakers.
For the test set, the target speaker spoke a keyword and then a command.
The command utterances are designed assuming a personal assistant system.
120 target utterances from 4 speakers and 120 interfering utterances from another 4 speakers were randomly mixed, and 10 different combination patterns resulted in a test set comprising 1,200 utterances.
The two speakers were located at different angles from the microphone array.
We call this test set a {\it simu-set}.

Another test set was also recorded in the same room, but the situation was more realistic.
The recording setting is presented in Fig.~\ref{fig:room}.
In this test set, one target speaker spoke under television sound.
The distance from the microphone array to the television was 1.2 meters, and that to the target speaker was also 1.2 meters.
The recording was performed by changing the speaker location (shown as A and B in the figure) and the volume of the television.
The number of utterances was 4,396 from 67 speakers.
We call this test set {\it real-set}.

\section{Signal-level evaluation}
\subsection{Evaluation metrics}
Firstly, the signal-level evaluation was performed to verify that the {\it keyword mask} can extract the keyword signal only from the mixture while the {\it non-keyword mask} can extract the remaining signal as well.
As evaluation measures, the signal-to-distortion ratio improvement (SDRi) \cite{Vincent} was used.
The SDRi represents the degree of reduction of the undesired signal and the extraction of the desired signal.
When this figure becomes positive, the estimated mask is thought to work properly.
Given a magnitude spectra of the desired signal $X_{t,f}$, that of the undesired signal $N_{t,f}$ and the mask $m_{t,f}$, the SDRi is calculated by:
%
%
\begin{eqnarray}
\textit{SDRi}=\frac{1}{F} \sum_{f \in {\bf F}} 10\log_{10}\Bigl({\frac{\sum_{\tau\in{\bf T}}{m_{\tau,f} {X}_{\tau,f} {X}_{\tau,f}^{*}} } {\sum_{\tau\in{\bf T}}{m_{\tau,f} {N}_{\tau,f}{N}_{\tau,f}^{*}}}}\Bigr) - {\xi},
\end{eqnarray}
where $f$, $F$ and ${\bf F}$ denote the frequency bin index, the total number of frequency bins, and the set of all frequency indexes respectively.
$\xi$ represents the SDR before masking and is defined as follows:
\begin{eqnarray}
\xi = \frac{1}{F} \sum_{f \in {\bf F}} 10\log_{10}\Bigl({\frac{\sum_{\tau\in{\bf T}}{ {X}_{\tau,f} {X}_{\tau,f}^{*}} } {\sum_{\tau\in{\bf T}}{{ {N}_{\tau,f}{N}_{\tau,f}^{*}}}}}\Bigr).
\end{eqnarray}

\subsection{Results}
The {\it simu-set} was used for evaluation.
In this work, the evaluation was performed using manually annotated keyword regions to separate the accuracies of keyword detection and the proposed method.
We compared the following four kinds of mask: estimated {\it keyword mask} $m^{(k)}$, estimated {\it non-keyword mask} $m^{(n)}$, IBM corresponding to the keyword $\textit{IBM}^{(k)}$, and IBM corresponding to the non-keyword $\textit{IBM}^{(n)}$.
Note that the keyword signal was regarded as $X_{t,f}$ for the evaluation of {\it keyword mask}, but it was regarded as $N_{t,f}$ for that of {\it non-keyword mask}.
Table \ref{tbl:signal} indicates the average and standard deviation of the SDRi, which was calculated from whole utterances of the {\it simu-set}.
We found from this table that the trained DNN-mask worked preferably, because the results of the two estimated masks were all positive.

\subsection{Example of the processing result}
\begin{figure}[tb]
\begin{center}
\vspace{-10pt}   
\includegraphics[width=8.0cm]{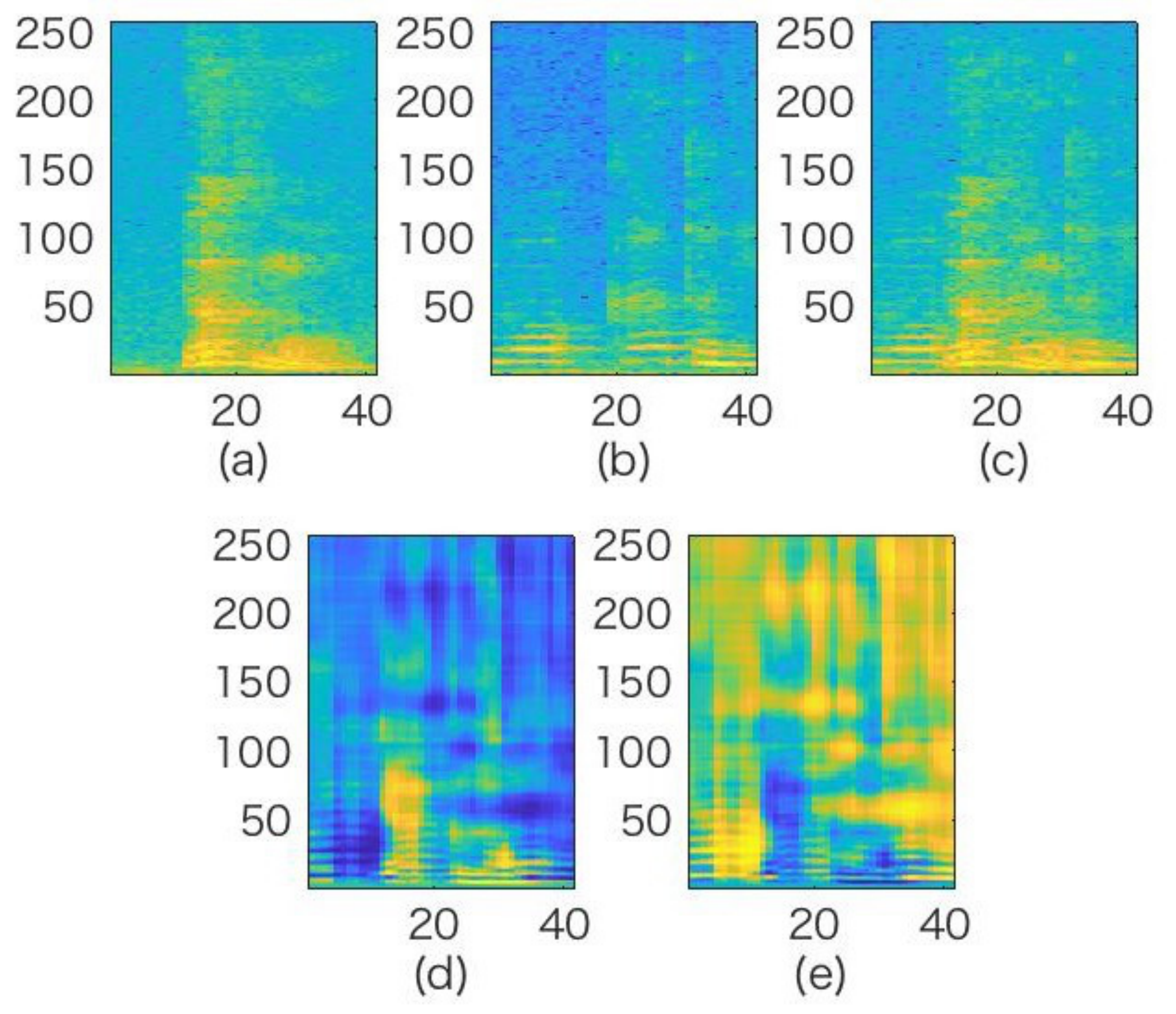}
\vspace{-15pt}     
\end{center}
\caption{An example of the DNN-based mask estimation result. (a): keyword speech, (b): background speech, (c): mixed speech, (d) estimated keyword mask, (e) estimated non-keyword mask.}
\label{fig:example-kwd}
\end{figure}

An example of the DNN-based mask estimation is presented in Fig.~\ref{fig:example-kwd}, which shows the spectrograms of the original keyword speech, background speech, and their mixed speech signal.
The estimated {\it keyword mask} and {\it non-keyword mask} are also presented.
From the figure, we can see that the estimated two masks selectively extracted the keyword and background speech respectively.
From this, it turned out that the trained DNN was able to model the desired behavior.
However, Fig.~\ref{fig:example-kwd} (d) also shows that the high frequency part of the keyword signal was erroneously missing from the {\it keyword mask} (see around frame 20).
We speculate that this was the reason for the performance gap of SNRi between $m^{(k)}$ and $\textit{IBM}^{(k)}$ seen in Table \ref{tbl:signal}.

\begin{figure}[tb]
\begin{center}
\vspace{-10pt}   
\includegraphics[width=7.1cm]{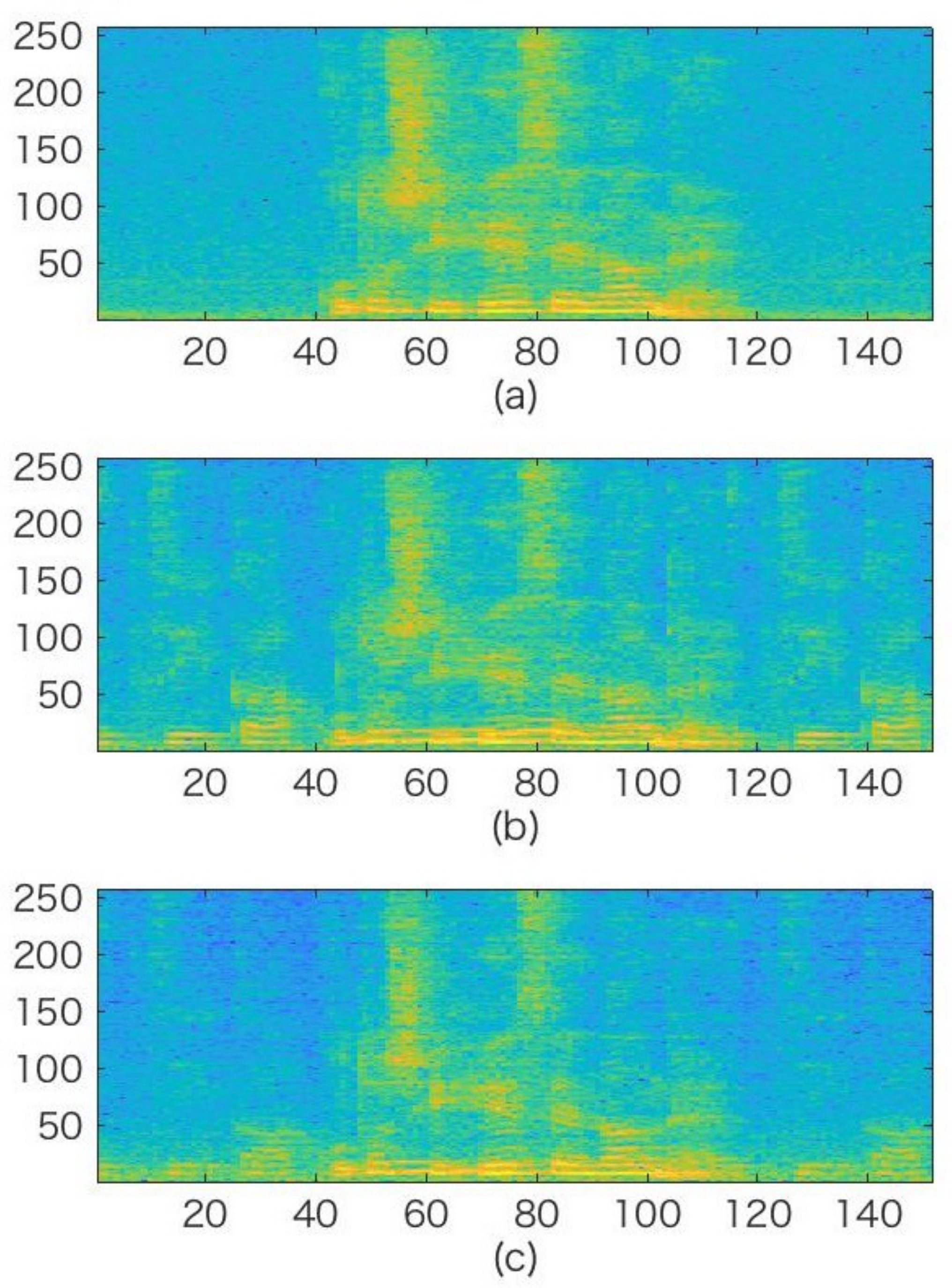}
\vspace{-10pt}     
\end{center}
\caption{An example of the proposed method. (a): command signal from a target speaker, (b): mixed speech, (c): processed mixed speech by the proposed method.}
\label{fig:example-cmd}
\vspace{-10pt}   
\end{figure}

Furthermore, the beamforming result using the estimated {\it keyword mask} and {\it non-keyword mask} shown in Fig.~\ref{fig:example-kwd} (d), (e) is also presented.
Fig.~\ref{fig:example-cmd} (a), (b) show spectrograms of a command speech from the target speaker and its mixed signal with background speech, which are subsequent signals seen in Fig.~\ref{fig:example-kwd} (a), (c).
Note that the shown spectrograms are one of the four recorded channels.
The beamforming result is also shown in Fig.~\ref{fig:example-cmd} (c).
From the figure, background speech was significantly reduced by the beamforming.
Therefore, our proposed method works well for this example.

\begin{table}
\caption{Comparison of SDRi between the estimated DNN-based mask and IBM. The table indicates average ($\mu$) and standard deviation ($\sigma$) as $\mu\pm\sigma$ in decibel (dB).}
\label{tbl:signal}
\begin{center}
\begin{tabular}{|l||l|l||l|l|}\hline
     & $m^{(k)}$ & $\textit{IBM}^{(k)}$ & $m^{(n)}$ & $\textit{IBM}^{(n)}$\\ \hline
SDRi &  4.9$\pm$2.6     & 11.1$\pm$3.4 & 3.1$\pm$1.6 & 10.8$\pm$4.0\\ \hline
\end{tabular}
\vspace{-10pt}   
\end{center}
\end{table}

\section{ASR evaluation}
\subsection{ASR system}
Next, the effectiveness of the beamforming using the trained DNN-based masks was verified with ASR experiments.
Our ASR system used a DNN-HMM (Deep Neural Network--Hidden Markov Model) based acoustic model.
The DNN had 5 fully connected hidden layers and each layer had 1,024 nodes.
The model parameters were trained with the cross entropy error criterion.
1,800 hours of speech data was used for training the acoustic model, which was collected through our voice service including search, dialogue and car-navigation.
The training utterances were split into three subsets, and 20\% of them were used directly.
40\% was mixed with various kinds of daily life noise, and various reverberation filters generated by a room simulator were added to the remaining 40\%.
The language model was a tri-gram model trained using text queries of the Yahoo!\ JAPAN search engine and transcriptions of mobile voice search queries.
The vocabulary size was about 1.6 million words.
Our decoder was an internally developed single-pass WFST decoder \cite{Iso}.
Prior to recognition, DNN-based voice activity detection (VAD) similar to \cite{Zhang} was performed to minimize insertion errors.

\subsection{Results for simu-set}
The results for the {\it simu-set} are firstly presented.
We compared the proposed method with some reference signals, and Table \ref{tbl:asr-simu} summarizes their results.
The table shows the character error rates (CER) and the relative error reduction rates (RERR) from the result when the mixed signal is input directly to the ASR, which is shown as `Mixed'.
'Clean' indicates the result of the target signal before mixing.
`Proposed' indicates the proposed method.
`Oracle (IBM)' indicates the result of oracle experiment that runs the proposed method with IBM calculated during the keyword region instead of the estimated masks.
Therefore, the result of `Oracle (IBM)' is thought to be an upper limit of the proposed method.
In addition, 'BeamformIt' shows the result of BeamformIt, a well known beamforming method \cite{Anguera}.

We would like to start our discussion by comparing `Clean' and `Mixed'.
From the table, we can see the error rates were drastically increased by mixing background speech.
`BeamformIt' showed it improved the error rate from `Mixed', but the improvement was not significant.
However, this result had been expected because BeamformIt simply estimated the beamforming filter from the observed signal, and it seemed difficult to selectively extract target speech from the mixture.
On the other hand, we can see that `Proposed' improved the CER significantly, even if the beamforming filter was estimated during only the short keyword utterance.
From this result, we confirmed the effectiveness of the proposed method for ASR under background speech.
However, the CER of `Oracle (IBM)' was smaller than `Proposed', and the performance gap was not trivial.
This means that there is still room to further improve our mask estimator.
\begin{table}
\caption{ASR result for simu-set. The proposed method shows significant CER improvement.}
\label{tbl:asr-simu}
\begin{center}
\vspace{10pt}   
\begin{tabular}{|l|ll|}\hline
                  & CER (\%) & RERR (\%)\\ \hline
Mixed             & 30.0 & - \\
BeamformIt        & 28.6 & 5.3 \\
Proposed          & 22.0 & 26.7 \\
Oracle (IBM)    & 18.6 & 38.0 \\
Clean             & 10.1 & 66.3 \\ \hline
\end{tabular}
\vspace{-10pt}   
\end{center}
\end{table}
\begin{table*}
\caption{ASR result for the real-set. The figures indicates CER (\%), and those in the brackets show RERR (\%) from the `Mixed'. Different from BeamformIt, the proposed method shows improvement in any conditions.}
\label{tbl:asr-real}
\begin{center}
\vspace{10pt}   
\begin{tabular}{|l|l||l|l|l|}\hline
Speaker position & TV-volume & Mixed & BeamformIt & Proposed\\ \hline
A            & Medium & 34.3     & 30.8 (10.2)    & 26.2 (23.6)\\
A            & Large  & 48.9     & 48.2 (1.4)     & 37.9 (22.5)\\
B            & Medium & 26.7     & 29.4 (-10.1)   & 24.1 (9.7)\\
B            & Large  & 38.1     & 43.4 (-13.9)   & 32.5 (14.7)\\ \hline
\end{tabular}
\vspace{-10pt}   
\end{center}
\end{table*}

\subsection{Results for real-set}
Table \ref{tbl:asr-real} shows the results for the real-set.
As described in Section \ref{sec:eval-data}, the evaluation was performed for four different acoustic conditions including two speaker locations and two levels of the television volume (medium, large).
Note that the results of IBM are not presented in this table as IBM could not be calculated from real recorded data.
The table shows that the CER was significantly decreased by our proposed method in all acoustic conditions.
Thus, the effectiveness of the proposed method could also be confirmed in the realistic situation.

\section{Conclusion}
This paper describes one solution for robust ASR under background speech.
The novelty of our approach is that it utilizes the wakeup keyword utterance in order to estimate spatial characteristics of the target speaker.
The proposed method firstly separated the mixture signal into the keyword and the remaining background speech using a DNN-based mask estimator, and then the separated signals were used for calculating a beamforming filter to enhance the subsequent utterances from the target speaker.
The signal-level evaluation showed that our DNN-based mask estimator could selectively separate these signals, and the effectiveness of the proposed method was also confirmed with ASR experiments.

Our future work includes the improvement of the mask estimator using a more sophisticated neural network architecture with more training data.
The verification of the proposed method under various noise conditions are also included in the future work.

\bibliographystyle{IEEEbib}
\bibliography{refs}

\end{document}